\documentclass[review]{elsarticle} 

\usepackage{amssymb}
\usepackage{amsfonts, amsmath, amssymb}
\usepackage{amsthm}
\usepackage{graphicx}
\usepackage{float}
\usepackage{enumitem}

\usepackage[utf8]{inputenc} 
\usepackage{enumitem}
\usepackage{lipsum}  

\usepackage{graphicx}
\usepackage{subcaption}
\usepackage[utf8]{inputenc}
\usepackage{url}
\DeclareMathOperator{\Tr}{Tr}

\begin{document}

\begin{frontmatter}



\title{Preventing epidemics by wearing masks: An application to COVID-19}


\author{João A. M. Gondim}

\address{Unidade Acad\^{e}mica do Cabo de Santo Agostinho, Universidade Federal Rural de Pernambuco, Cabo de Santo Agostinho, PE, Brazil}

\begin{abstract}
The goal of this work is to consider widespread use of face masks as a non-pharmaceutical control strategy for the Covid-19 pandemic. A SEIR model that divides the population into individuals that wear masks and those that do not is considered. After calculating the basic reproductive number by a next generation approach, a criterion for determining when an epidemic can be prevented by the use of masks only and the critical percentage of mask users for disease prevention in the population are derived. The results are then applied to real world data from the United States, Brazil and Italy.

\end{abstract}




\end{frontmatter}


\section{Introduction}
\label{}

The Covid-19 crisis has created the biggest public health concerns of 2020. Since being first reported at the end of 2019, the disease has caused over 23 million confirmed cases and 800 thousand deaths by the end of August 2020 \cite{who}. Many studies to model the pandemic spread were developed (e.g. \cite{ferguson2020report,wu2020nowcasting,li2020substantial,ngonghala2020mathematical}) with results influencing the policies of governments around the world.

Attempts to minimize the damage of the pandemic were then implemented, such as mandatory mask use and quarantines, which improved the overall scenario, but so far there are no drugs or vaccines to treat or immunize people and maintaining quarantines for longer time periods is not a viable option in some communities. Therefore, looking for non-pharmaceutical control strategies is essential to deal with this epidemic and others in the future.

This paper addresses this issue considering the widespread use of masks as in other works such as \cite{eikenberry2020mask,chan2020covid,cheng2020role,li2020mask}. A SEIR model \cite{li1999global} with individuals divided into those that wear masks and those that do not is considered in Section 2, and its basic reproductive number is calculated by a next generation approach in Section 3. This leads to a criterion that determines when an epidemic outbreak can be avoided by mask use only, and a critical percentage of mask users in the population is derived. 

For applications of these methods, we perform the parameter fitting in Section 4 with data from the United States, Brazil and Italy and analyse whether the Covid-19 crisis could have been avoided in these countries with widespread mask use from the beginning of the outbreak. In addition, numerical simulations are carried out to verify how the evolution of the disease is mitigated if it cannot be avoided. The conclusions are drawn in Section 5.

\section{Model structure}
\label{}

Consider a population $N$ that is divided into individuals that wear masks, denoted by $N_m$ and individuals that do not, denoted by $N_n$. Let $p(t)$ be the percentage of people wearing a mask in the population at time $t \geq 0$, then
\begin{equation}
    N_m(t) = p(t)N(t), \quad N_n(t) = (1-p(t))N(t).
    \label{percs}
\end{equation}

Both $N_n$ and $N_m$ are also divided into four epidemiological classes, consisting of susceptible, exposed, infected and removed individuals, denoted by $S_n$ and $S_m$, $E_n$ and $E_m$, $I_n$ and $I_m$ and $R_n$ and $R_m$, respectively. As the model will consider only a short time period in comparison to the demographic time frame, vital parameters will be neglected, so the total population will be assumed constant, that is, 
\begin{equation}
    N(t) = N.
    \label{N-const}
\end{equation}

Let $r$ be a multiplicative factor for the transmission rate $\beta$ that will take into account the reduction in the probability of contagion from one person wearing a mask in a susceptible-infected contact. We assume that this reduction is the same whether a susceptible or an infective is wearing the mask. When only one individual has a mask on, we assume that the new transmission rate is $r\beta$. In the case of both individuals with masks on, then the transmission rate is assumed to be $r^2\beta$. There are four ways contagions can occur, and they are described in Table \ref{transmission}.

\begin{table}[!h]
 	\centering
 	\caption{Transmission possibilities.}
 	\label{transmission}
 	\begin{tabular}{ccc}
 		\hline 
 		\\
 		Susceptible & Infected  &   Transmission term    \\ 
 		\\
 		 \hline
\\
 		$S_n$ & $I_n$ & $\displaystyle{\frac{\beta S_n I_n}{N}}$ \\ \\
 		$S_n$ & $I_m$ & $\displaystyle{\frac{r\beta S_n I_m}{N}}$ \\ \\
 		$S_m$ & $I_n$ & $\displaystyle{\frac{r\beta S_m I_n}{N}}$ \\ \\
 		$S_m$ & $I_m$ & $\displaystyle{\frac{r^2\beta S_m I_m}{N}}$ \\ \\
 		 \hline
 	\end{tabular}%
\end{table}

We further assume that $p(t) = p$ is constant, so $N_m(t)$ and $N_n(t)$ are also constant. Hence, our model can be written as




\begin{equation}
    \begin{aligned}
        S_n' &= -\frac{\beta S_n}{N}\left(I_n +rI_m\right) \ , \\[0.4cm]
        S_m' &= -\frac{r \beta S_m}{N}\left(I_n +rI_m\right) \ , \\[0.4cm]
        E_n' &= \frac{\beta S_n}{N}\left(I_n +rI_m\right) - \sigma E_n \ , \\[0.4cm]
        E_m' &= \frac{r \beta S_m}{N}\left(I_n +rI_m\right) - \sigma E_m \ , \\[0.4cm]
        I_n' &= \sigma E_n - \gamma I_n \ , \\[0.4cm]
        I_m' &= \sigma E_m - \gamma I_m \ , \\[0.4cm]
        R_n' &= \gamma I_n \ , \\[0.4cm]
        R_m' &= \gamma I_m. \\[0.4cm]
    \end{aligned}
    \label{model1}
\end{equation}

The parameters $\sigma$ and $\gamma$ denote the exit rates from the exposed and infected classes, respectively. It is typically assumed that $\sigma = 1/T_e$ and $\gamma = 1/T_i$, where $T_e$ and $T_i$ are the mean lengths of the latency and infectious periods, respectively. 


\section{The basic reproductive number and some consequences}
\label{}

Now, the basic reproductive number, $R_0$, of model \eqref{model1} is calculated. This will be done by a next generation approach (see \cite{martcheva2015introduction,diekmann1990definition}). $R_0$ is given by the spectral radius of $$K = FV^{-1},$$ where
$$F=\left( \begin{array}{cccc} 0 & 0 & \beta(1-p) & \beta r(1-p) \\ 0 & 0 & \beta rp & \beta r^2 p \\ 0 & 0 & 0 & 0 \\ 0 & 0 & 0 & 0 \\ \end{array} \right)$$ and $$V = \left( \begin{array}{cccc} \sigma & 0 & 0 & 0 \\ 0 & \sigma & 0 & 0 \\ -\sigma & 0 & \gamma & 0 \\ 0 & -\sigma & 0 & \gamma \\ \end{array} \right).$$

Hence, $K = FV^{-1}$ is $$K = \left( \begin{array}{cccc} \frac{\beta(1-p)}{\gamma} & \frac{r\beta(1-p)}{\gamma} & \frac{\beta(1-p)}{\gamma} & \frac{r\beta(1-p)}{\gamma} \\ \frac{r\beta p}{\gamma} & \frac{r^2 \beta p}{\gamma} & \frac{r \beta p}{\gamma} & \frac{r^2 \beta p}{\gamma} \\ 0 & 0 & 0 & 0 \\ 0 & 0 & 0 & 0 \\ \end{array} \right).$$

Due to its block structure, the eigenvalues of $K$ are exactly the eigenvalues of $$K_{11} = \left( \begin{array}{cc} \frac{\beta(1-p)}{\gamma} & \frac{r\beta(1-p)}{\gamma} \\ \frac{r\beta p}{\gamma} & \frac{r^2 \beta p}{\gamma} \\ \end{array} \right).$$

It is clearly seen that the trace and the determinant of $K_{11}$ are $$\Tr(K_{11}) = \frac{\beta}{\gamma}\left(1-p+r^2p\right), \quad \det(K_{11}) = 0,$$ respectively, hence its eigenvalues are $0$ and 
\begin{equation}
    R_0 = \mathcal{R}_0\left[1-p\left(1-r^2\right)\right],
    \label{R0}
\end{equation}
where 
\begin{equation}
    \mathcal{R}_0 = \frac{\beta}{\gamma},
    \label{R0-SEIR}
\end{equation}
which is the basic reproductive number for the standard SEIR model without vital dynamics \eqref{modelSEIR} \cite{hethcote2009basic}.
\begin{equation}
    \begin{aligned}
        S' &= -\beta \frac{SI}{N} \ , \\[0.4cm]
        E' &= \beta\frac{SI}{N} - \sigma E \ , \\[0.4cm]
        I' &= \sigma E - \gamma I \ , \\[0.4cm]
        R' &= \gamma I. \\
        \label{modelSEIR}
    \end{aligned}
\end{equation}
Note that \eqref{model1} reduces to \eqref{modelSEIR} if either $p = 0$ (nobody wears masks) or $r = 1$ (the masks offer no protection against the disease).

\begin{figure}[h!]
	\centering
	\includegraphics[scale=0.7,trim={3.5cm 17.0cm 11.5cm 2.0cm}]{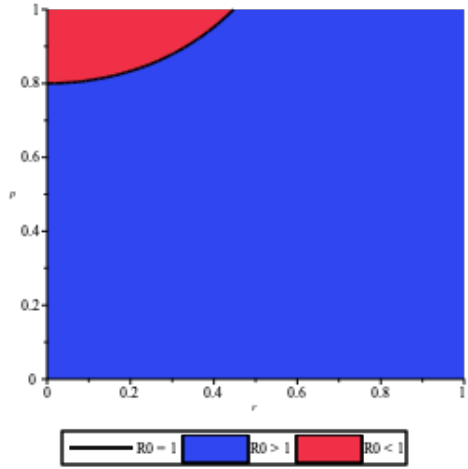}
	\caption{The level set of $R_0 = 1$ for $\mathcal{R}_0 = 5$ in the $rp$ plane.}
	\label{level} 
\end{figure}

Since both $p, r^2 \in [0,1]$, it is clear that $R_0 > 0$ and that $R_0 \leq \mathcal{R}_0$, with the equality in the cases $p = 0$ or $r = 1$, when the model \eqref{model1} reduces to model \eqref{modelSEIR}. The level set $R_0 = 1$ in \eqref{R0} is displayed in Figure \ref{level} for $\mathcal{R}_0 = 5$, along with the regions of $R_0 < 1$ and $R_0 > 1$. 

\begin{figure}[h!]
	\centering
	\includegraphics[scale=0.5,trim={3.5cm 0.5cm 3.5cm 0.5cm}]{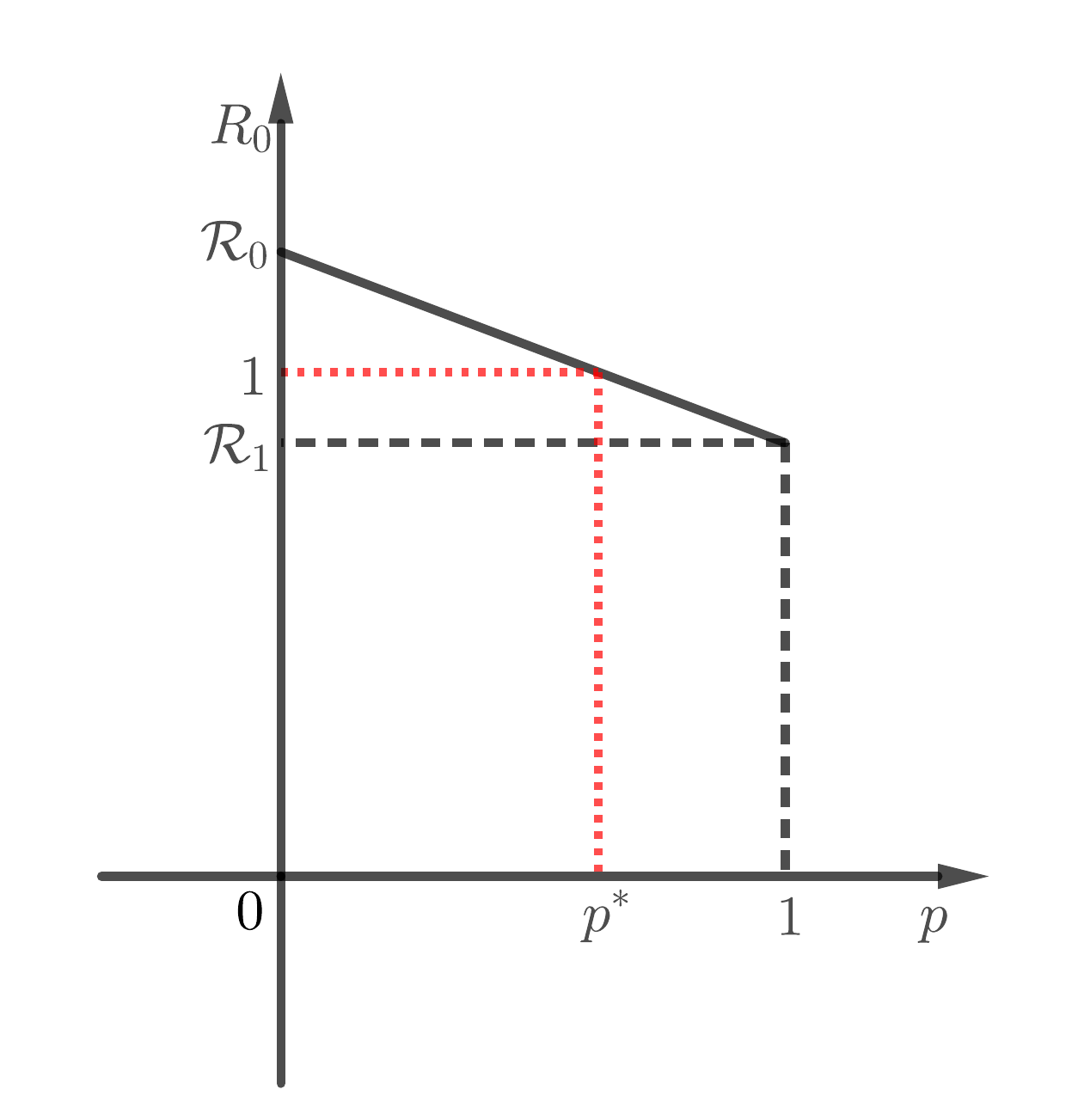}
	\caption{Plot of $R_0$ as a function of $p$.}
	\label{Fig1} 
\end{figure}

We now look for conditions the pair $(r,p)$ should satisfy in order for this point to lie in the region of $R_0 < 1$. Notice that, for fixed $r \in [0,1]$, $R_0$ is a decreasing function of $p$ (see Figure \ref{Fig1}). For $p = 0$, we have $R_0 = \mathcal{R}_0$, and for $p = 1$, we have $R_0 = \mathcal{R}_1$, where 
\begin{equation}
    \mathcal{R}_1 = \mathcal{R}_0 \cdot  r^2.
    \label{R1}
\end{equation}

It is assumed that $\mathcal{R}_0 > 1$. Then, it is clear from Figure \ref{Fig1} that one can find values of $p$ such that $R_0 < 1$ if and only if $\mathcal{R}_1 < 1$, that is, if and only if 
\begin{equation}
    r < \frac{1}{\sqrt{\mathcal{R}_0}}.
    \label{r-cond}
\end{equation}

Moreover, there is a critical value $p^*$ such that $R_0(p^*) = 1$, so $R_0(p) < 1$ if and only if $p > p^*$. Solving $R_0 = 1$ in \eqref{R0}, one sees that 
\begin{equation}
    p^* = \frac{1}{1-r^2}\left(1-\frac{1}{\mathcal{R}_0}\right).
    \label{crit_p}
\end{equation}

 The value of $p^*$ corresponds to the critical percentage of the population that should wear masks in order to avoid the epidemic outbreak. In the extreme case of $r=0$, i.e., the masks are ideal and avoid contamination for users, which is the same as immunizing the population, \eqref{crit_p} becomes
\begin{equation}
    p^* = 1-\frac{1}{\mathcal{R}_0},
    \label{herd-imm}
\end{equation}
which coincides with the usual threshold for herd immunity \cite{fine2011herd}.

\section{Data fitting and numerical results}
\label{}

In this Section, we collect data from the United States, Brazil and Italy to use as case studies for the results of the previous Section. The time frame in consideration consists of the first $30$ days after the cumulative number of cases in each country reached $100$, which happened in March 2, March 13 and February 23 for the US, Brazil and Italy, respectively. The data, which was retrieved from \cite{worldometers}, is displayed in Tables \ref{data-US}, \ref{data-BRA} and \ref{data-ITA}.

\begin{table}[!h]
 	\centering
 	\small
 	\caption{Cumulative cases in the USA starting at the first day with more at least 100 cases.}
 	\label{data-US}
 	\begin{tabular}{cccccccccc}
 		\hline 
 		
 		 Day & Cases & Day & Cases & Day & Cases & Day & Cases & Day & Cases   \\ 
 		
 		 \hline
 		      1 & 100 & 7 & 541    & 13 & 2,774  & 19 & 19,608 & 25 & 86,668  \\
              2 & 124 & 8 & 704    & 14 & 3,622  & 20 & 24,498 & 26 & 105,584 \\
              3 & 158 & 9 & 994    & 15 & 4,611  & 21 & 33,946 & 27 & 125,250 \\
              4 & 221 & 10 & 1,301 & 16 & 6,366  & 22 & 44,325 & 28 & 145,526 \\
              5 & 319 & 11 & 1,631 & 17 & 9,333  & 23 & 55,579 & 29 & 168,835 \\
              6 & 435 & 12 & 2,185 & 18 & 13,935 & 24 & 69,136 & 30 & 194,127 \\
            \hline
     	\end{tabular}%
        \end{table}

    \begin{table}[!h]
 	\centering
 	\small
 	\caption{Cumulative cases in Brazil starting at the first day with at least 100 cases.}
 	\label{data-BRA}
 	\begin{tabular}{cccccccccc}
 		\hline 
 		Day & Cases & Day & Cases & Day & Cases & Day & Cases & Day & Cases   \\ 
 		
 		 \hline
             1 & 151 &  7 & 640   & 13 & 2,554 & 19 & 5,717  & 25 & 12,183 \\
             2 & 151 &  8 & 970   & 14 & 2,985 & 20 & 6,880  & 26 & 14,034 \\
             3 & 200 &  9 & 1,178 & 15 & 3,417 & 21 & 8,044  & 27 & 16,188 \\
             4 & 234 & 10 & 1,546 & 16 & 3,904 & 22 & 9,194  & 28 & 18,145 \\
             5 & 346 & 11 & 1,924 & 17 & 4,256 & 23 & 10,360 & 29 & 19,789 \\
             6 & 529 & 12 & 2,247 & 18 & 4,630 & 24 & 11,254 & 30 & 20,962 \\
            
        \hline
        \end{tabular}%
        \end{table}

    \begin{table}[!h]
 	\centering
 	\small
 	\caption{Cumulative cases in Italy starting at the first day with at least 100 cases.}
 	\label{data-ITA}
 	\begin{tabular}{cccccccccc}
 		\hline 
 		Day & Cases & Day & Cases & Day & Cases & Day & Cases & Day & Cases   \\ 
 		
 		 \hline
             1 & 157 & 7  & 1,128 & 13 & 4,639  & 19 & 15,122 & 25 & 35,732 \\
             2 & 229 & 8  & 1,702 & 14 & 5,886  & 20 & 17,670 & 26 & 41,056 \\
             3 & 323 & 9  & 2,038 & 15 & 7,380  & 21 & 21,169 & 27 & 47,044 \\
             4 & 470 & 10 & 2,504 & 16 & 9,179  & 22 & 24,762 & 28 & 53,598 \\
             5 & 655 & 11 & 3,092 & 17 & 10,156 & 23 & 27,997 & 29 & 59,158 \\
             6 & 889 & 12 & 3,861 & 18 & 12,469 & 24 & 31,524 & 30 & 63,941 \\
        
         		    \hline
 	\end{tabular}%
\end{table}

We assume that the mean latency and recovery periods are $5.1$ and $7$ days, respectively, as in \cite{eikenberry2020mask}. We fit $\beta$ by a minimization routine based on the least squares method, available in \cite{martcheva2015introduction}, in the standard SEIR model without vital dynamics \eqref{modelSEIR}.  The routine minimizes the difference of the cumulative number of cases, given by $I(t)+R(t)$, and the data points.

The total populations of the USA, Brazil and Italy will be rounded to $331$, $209$ and $60$ million, respectively. These numbers will be taken as the initial values of susceptible individuals in each country. In the first days of each data set, the numbers of active cases (see \cite{worldometers}) were $85$, $150$ and $152$ for the USA, Brazil and Italy, respective, so the initial conditions for infected and removed individuals will be taken, respectively, as $85$ and $15$ for the USA, $150$ and $1$ for Brazil and $152$ and $5$ for Italy.

For the initial values of exposed individuals, we use the fact that the latency period is taken as $5.1$ days, so we look at the number in day $6$ of each data set and choose the initial number of exposed as the extra number of cases since day $1$. Hence, these numbers are $335$ in the USA, $378$ in Brazil and $732$ in Italy. A summary of the initial conditions for each country in the minimization routine is shown in Table \ref{summary}.

\begin{table}[!h]
 	\centering
 	\small
 	\caption{Initial conditions for the estimation of $\beta$.}
 	\label{summary}
 	\begin{tabular}{ccccc}
 		\hline 
 		Country & S(0) & E(0) & I(0) & R(0)   \\ 
 		
 		 \hline
             USA & 331 million & 335 & 85 & 15 \\
             Brazil & 209 million & 378 & 150 & 1 \\
             Italy & 60 million & 732 & 152 & 5 \\
        
         		    \hline
 	\end{tabular}%
\end{table}

Starting with an initial guess of $\beta = 0.5$, the fitted values of $\beta$ are, then, 
\begin{equation}
    \beta_{US} = 0.8577, \quad \beta_{BR} = 0.4854, \quad \beta_{IT} = 0.5809.
    \label{betas}
\end{equation}

Using \eqref{R0-SEIR}, we can calculate its value for each country. The results are displayed in Table \ref{R0-values}.

\begin{table}[!h]
 	\centering
 	\small
 	\caption{Basic reproductive number for each country in the standard SEIR model.}
 	\label{R0-values}
 	\begin{tabular}{cc}
 		\hline 
 		Country & $\mathcal{R}_0$ \\ 
 		
 		 \hline
             USA & $6.0039$ \\
             Brazil & $3.3978$ \\
             Italy & $4.0663$ \\
        
         		    \hline
 	\end{tabular}%
\end{table}


According to \cite{chu2020physical}, when both individuals wear masks in a susceptible-infected contact, there is an average reduction of $82.18\%$ in the transmission, so $$r^2 = 1 - 0.8218 = 0.1782,$$ hence $r = 0.4221$. Then, the result of one person wearing a mask when two individuals meet is a decrease of around $58\%$ in the transmission coefficient.

Rewriting \eqref{r-cond}, we see that the outbreak can be avoided by a widespread use of masks is $$\mathcal{R}_0 < \frac{1}{r^2} \approx 5.6117.$$ According to Table \ref{R0-values}, the Covid-19 crisis could have been avoided in Brazil and in Italy, and according to \eqref{crit_p}, this would be possible if at least $85.87\%$ and $91.76\%$ of all individuals wore masks, respectively.

On the other hand, Table \ref{R0-values} indicates that Covid-19 could not have been avoided in the USA by widespread mask use only, but the basic reproductive number could be lowered from $\mathcal{R}_0 = 6.0039$ to $\mathcal{R}_1 = 1.0699$, so other a combination with control measures such as social distancing, quarantines and tracking infected cases could be able to prevent the disease outbreak. This could also be done by improving the average mask protection. 

Moreover, one could also calculate the normalized forward sensitivity index (or elasticity) of $R_0$ with respect to $p$ (see \cite{martcheva2015introduction,chitnis2008determining}), given by
\begin{equation}
    \Upsilon_p^{R_0} = \frac{\partial R_0}{\partial p}\cdot \frac{p}{R_0}, 
    \label{index}
\end{equation}

This number provides the percentage change in $R_0$ for a given percentage change in $p$. For example, if $\Upsilon_p^{R_0} = -0.5$, then a $10\%$ increase in $p$ produces a $5\%$ decrease in $R_0$. By \eqref{R0} and \eqref{index}, we have
\begin{equation}
    \Upsilon_p^{R_0} = -\frac{p(1-r^2)}{1-p(1-r^2)}.
    \label{index-p}
\end{equation}

\begin{figure}[ht]
	\centering
	\includegraphics[scale=0.75,trim={5.5cm 10.0cm 5.5cm 10.0cm}]{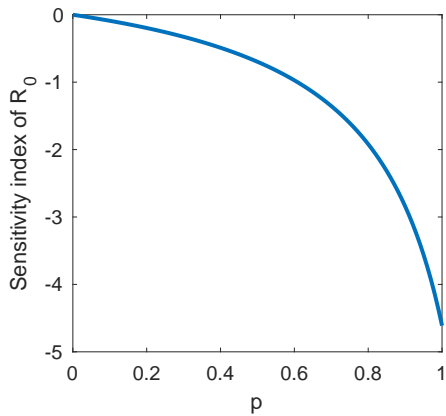}
	\caption{Plot of the normalized forward sensitivity index of $R_0$ with respect to $p$ as a function of $p$.}
	\label{Fig2} 
\end{figure}

A plot of $\Upsilon_p^{R_0}$ as a function of $p$ is displayed in Figure \ref{Fig2}. It shows that $R_0$ becomes very sensitive to $p$ for bigger values of this parameter, so even if most of a community has already become adept to wearing masks, small increases in $p$ could still contribute greatly to epidemic control.

\begin{figure}[ht]
	\centering
	\includegraphics[scale=0.7,trim={5.5cm 8.0cm 5.5cm 9.5cm}]{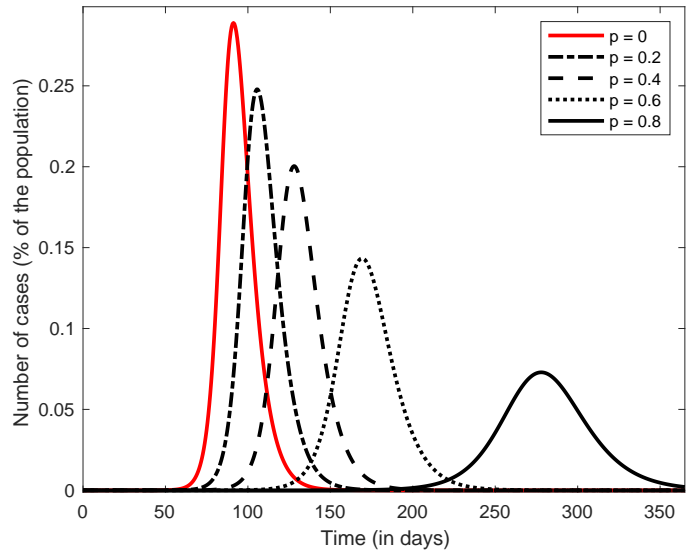}
	\caption{Plots of the infected curves for different values of $p$ in model \eqref{model1} in the case of the USA.}
	\label{Fig3} 
\end{figure}

We now assess the effect of variations in $p$ on the numbers of infected caused by Covid-19 in the case of the USA. A comparison of the total infected curve $$I_n(t)+I_m(t),$$ normalized by the total population and with initial conditions 
\begin{equation}
\begin{aligned}
    S_n(0) &= (1-p)N, \quad  &S_m(0) &= pN, \quad  &E_n(0) &= 0, \quad &E_m(0) &= 0, \\ I_n(0) &= 0, \quad  &I_m(0) &= 1, \quad  &R_n(0) &= 0, \quad &R_m(0) &= 0, \\
\end{aligned}
    \label{ic}
\end{equation}
where $N = 331$ million, is displayed in Figure \ref{Fig3} for a time period of one year.

\begin{figure}[h!]
	\centering
	\includegraphics[scale=0.8,trim={5.5cm 10.5cm 5.5cm 9.5cm}]{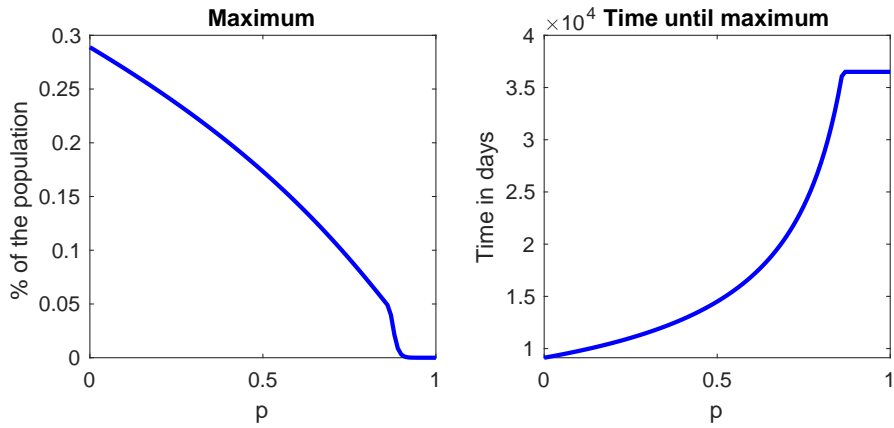}
	\caption{Plots of the infected curves for different values of $p$ in model \eqref{model1}.}
	\label{Fig4} 
\end{figure}

The desired ''flattening of the curve``, i.e., postponing and lowering the maximum number of cases, is achieved. A closer look at this fact is shown in Figure \ref{Fig4}, which shows that, in a period of one year, both the maximum and the time it happens stabilize after $p \approx 0.86$. For these values of $p$, plots like the ones in Figure \ref{Fig3} would only reach their peak after one year, so we can say that the disease is essentially controlled.

\section{Conclusion}
\label{}

In this paper, a SEIR model is considered in a population that is divided into individuals that wear masks and individuals that do not. Parameters $p$ and $r$, which represent the (constant) percentage of the population that are mask users and the reduction in the transmission rate due to one person wearing a mask in a susceptible-infected contact, respectively, are introduced, and their effect on the basic reproductive number is calculated by a next generation method.

This allows for the derivation of a necessary and sufficient condition for epidemic outbreaks to be prevented only by the widespread use of masks. When this is possible, a critical percentage $p^*$ of mask users in the population necessary for disease control is calculated. 

This is utterly important in dealing with public health crisis worldwide, since pharmaceutical measures such as vaccines and drugs are more laborious and take long times to be developed while diseases spread.

As case studies for the results in this paper, real world data from the Covid-19 pandemic was used, focusing on the United States, Brazil and Italy for the first 30 days after the total number of cases reached 100. After fitting the parameters, the results implied that the Covid-19 epidemic could have been avoided in Brazil and Italy if at least 85.87\% and 91.76\% of the populations, respectively, wore masks from the beginning of the outbreak.

Even though this was not possible in the case of the United States, we noted that the basic reproductive number could have been reduced from 6.0039 to 1.0699, so other control measures such as social distancing, quarantines, or even improving the average mask quality could help pushing this number below 1.

Furthermore, numerical simulations showed that the flattening of the infected curve is achieved as $p$ gets closer to 1, and that the maximum of this curve and the necessary time for it to happen stabilize after $p \approx 0.86$, i.e., the disease is essentially controlled. Thus, simple measures such as wearing masks can prove to be very effective in controlling, or even preventing, future epidemics.




 \bibliographystyle{elsarticle-num} 
 \bibliography{references.bib}





\end{document}